\pdfoutput=1
\documentclass[fleqn,usenatbib]{mnras}  

\volume{{\rm ArXiV post-print, 2019/04/04}}

\usepackage{graphicx}	
\usepackage{amsmath}	
\usepackage{amssymb}	
\usepackage{array}
\usepackage{tikz}
\usetikzlibrary{arrows} 
\usetikzlibrary{shapes.multipart}
\usetikzlibrary{shapes.misc, positioning}
\usepackage{todonotes} 
\usepackage[export]{adjustbox}

\hypersetup{pdfauthor={N.A.Bharmal on behalf of O.B-M. & C.M.C.},
            pdftitle={Atmospheric profiling on seeing-limited image]{PEPITO: atmospheric Profiling from short-Exposure focal Plane Images in seeing-limiTed mOde}},
            pdfkeywords={atmosphere, turbulence, profile, optical turbulence, otp, atmospheric profile, fpp, psf model},
            bookmarksnumbered=true}

\newcommand{\para}[1]{\left(#1\right)}

\newcommand{\cro}[1]{\left[#1\right]}

\newcommand{\rz}{r_0}

\newcommand{\cnh}{C_n^2(h)}
\newcommand{\psf}[1]{\text{PSF}_{#1}}

\newcommand{\otf}[1]{\text{OTF}_{#1}}

\newcommand{\atf}[1]{\text{ATF}_{#1}}

\newcommand{\rhob}{\boldsymbol{\rho}}
\newcommand{\rbb}{{\boldsymbol{r}}} 
\newcommand{\pup}{\mathcal{P}}
\newcommand{\FTT}{\mathcal{F}_\text{TT}}
\newcommand{\josa}{JOSA } 
\newcolumntype{P}[1]{>{\centering\arraybackslash}p{#1}}

\title[Atmospheric profiling on seeing-limited image]{PEPITO: atmospheric Profiling from short-Exposure focal Plane Images in seeing-limiTed mOde}

\author[O. Beltramo-Martin et al.]{
	O. Beltramo-Martin$^{1,2}$\thanks{E-mail: olivier.beltramo-martin@lam.fr},
	N.A Bharmal$^{3}$,		
	C.M. Correia$^{1}$\\
	$^{1}$ONERA, The French Aerospace Laboratory  BP. 72, F-92322 Chatillon Cedex, France\\	
	$^{2}$Aix Marseille Univ., CNRS, CNES LAM, 38 rue F. Joliot-Curie, 13388 Marseille, France\\	
	$^{3}$Centre for Advanced Instrumentation (CfAI), Durham University, South Road, Durham, DH1 3LE, United Kingdom.}

\date{Accepted 2019 April 04. Received 2019 March 29; in original form 2019 January 29}
\pubyear{2010}

\begin{document} 
	\label{firstpage}
	\pagerange{\pageref{firstpage}--\pageref{lastpage}}
	\maketitle
	\begin{abstract}
		Atmospheric profiling is a requirement for controlling wide-field Adaptive Optics (AO) instruments, analyzing the AO performance with respect to the observing conditions and predicting the Point Spread Function (PSF) spatial variations. 	We present PEPITO, a new concept for profiling atmospheric turbulence from {\em post~facto} tip-tilt (TT) corrected short-exposure images. PEPITO utilizes the anisokinetism effect in the images between several stars separated from a reference star, and then produces the profile estimation using a model-fitting methodology, by fitting to the long exposure TT-corrected PSF.  PEPITO	has a high sensitivity to both $\cnh$ and $L_0(h)$ by relying on the full telescope aperture and a large field of view. It then obtains a high vertical resolution (1\,m-400\,m) configurable by the camera pixel scale, taking advantage of fast statistical convergence (of order of tens of seconds). With only a short exposure-capable large format detector and a numerical complexity independent of the telescope diameter, PEPITO perfectly suits accurate profiling for night optical turbulence site characterization or adaptive optics instruments operations. We demonstrate, in simulation, that the $\cnh$ and $L_0(h)$ can be estimated to better than 1\% accuracy, from fitted PSFs of magnitude V=11 on a D=0.5\,m telescope with a 10 arcmin field of view.
	\end{abstract}
	
		
\section{Introduction}
\label{S:Intro}

Atmospheric characterization for a ground-based telescope has become a key step in the design of instrumentation to correct for wave-front aberrations introduced by atmospheric turbulence. Adaptive Optics (AO) compensates the wave-front aberrations in real-time and the benefits gained from knowledge of the atmospheric profile, or $\cnh$, include operating the tomographic turbulence compensation in wide-field using multiple Guide Stars (GS) \citep{Ono2018,Ono2016,Correia2015,Vidal2010}, enabling phase predictive control \citep{Correia2017, Males2018,Juvenal2016,Sivo2015,Rudy2014,Petit2014} for optimum AO performance, or providing a comprehensive analysis of AO residuals \citep{Ferreira2018,Martin2017}. High altitude layers play a major role in the spatial phase decorrelation \citep{Roddier1981} and has an impact on wide-field AO performance \citep{Fusco2012} and extreme AO \citep{Cantalloube2018}. This calls for a high-accuracy, high-altitude profile identification technique.  

At present, profiling using AO instruments is performed from the cross-correlation of Wave-Front Sensor (WFS) measurements  \citep{Laidlaw2018,Mazzoni2016,Ono2017,Martin2016L3S,Neichel2014,Vidal2010}, with profile height limits of $\sim10$--20\,km imposed principally by the telescope diameter, and then additionally the cone effect for laser-based AO \citep{Foy2000}. However, such an approach is only available on multiple GS-based systems and can not be deployed to predict PSF variations on images delivered by single-conjugated AO systems. To achieve this prediction, we must rely on dedicated standalone profilers, such as Stereo-Scidar~\citep{Osborn2018}, MASS/DIMM~\citep{Butterley2018}, FASS~\cite{Guesalaga2016FASS} or SLODAR~\cite{Wilson2002} instruments, or use predictive weather model~\citep{Osborn2018_GCM,Masciadri2017}. 

For PSF modeling purpose, the relevant metric is the focal-plane image, calling for a new type of image-based technique that is capable of retrieving the atmospheric profile from the image itself and not from information of a different nature. In this context, we have proposed the Focal Plane Profiling (FPP) algorithm \citep{BeltramoMartin2018_FPP} to retrieve the atmospheric profile from the anisoplanatism-affected images through fitting PSF models \citep{BeltramoMartin2018_aniso} across various points in the field of view (FOV). The use of the focal-plane image is particularly relevant to calibrate the anisoplanatism model regarding the key metric that is the PSF, which can feed algorithms of deconvolution or model-fitting. This technique has revealed to be efficient but needs post-AO images of point sources to be operable, which limits its range of applicability regarding the presence of a sufficient number of bright stars in the field.  \\

To keep the strength of a FPP-like approach but make it independent to the AO system,  we propose in this paper a novel atmospheric profiling concept, named PEPITO, which uses on-axis Tip-tilt (TT) corrected focal-plane images. PEPITO relies on the TT-anisoplanatism effect \citep{Fried1982} (commonly called anisokinetism) which elongates the off-axis Point Spread Function (PSF) with respect to the $\cnh$ profile. This anisoplanatism is created either digitally ({\em post~facto}) from short exposure images (1-10\,ms) or by a real-time compensation using a dedicated device. Therefore, we create anisokinetism-contaminated PSFs, without need of AO, which are passed to the FPP algorithm to characterize the atmosphere vertical distribution. \\

Whereas {PEPITO} operates from long-exposure focal plane images, therefore using the entire pupil as an aperture, cross-correlation methods instead utilize sub-apertures and then from across all either their centroids (SLODAR) or their scintillation (SCIDAR) is correlated. Consequently, for a given telescope aperture, PEPITO benefits from a full aperture gain \citep{Plantet2013} that provides a better signal-to-noise ratio (S/N) for retrieving atmosphere statistics ($\cnh\, L_0(h)$).

Each stellar image is spread across a number of detector pixels and there are two important characteristics to consider: the separation from the reference star (baseline) and the change in image (morphology). The baseline gives access to the decorrelation in angle and this is most sensitive for a certain altitude range. The presence of a turbulent layer elongates the PSF in the reference star direction, making the aspect ratio maximal for a baseline value that decrease with respect to the layer height. The morphology encodes the $\cnh$ and $L_0(h)$ for that altitude range. The range of baselines and the FOV then constrains the altitude limit. The cross-correlation methods have fundamentally a baseline-equivalent angular separation of two stars together with spacings of the sub-apertures in the pupil, from their widths to the diameter of the pupil. The angular separation and sub-aperture size or pupil diameter determines, respectively, the altitude spacing of the profile and the upper altitude limit. Therefore despite both methods reliance on the angular decorrelation of phase from atmospheric turbulence, and for {PEPITO} and SLODAR specifically the TT component, the resulting characteristics favour {PEPITO} for high-resolution profiling at better that few hundreds of meters.  Finally, PEPITO relies on long-exposure PSFs that have statistically converged, i.e. their aspect ratio and Full Width at Half Maximum (FWHM) does not vary by accumulating more frames, \textcolor{black}{which can be reached in tens of seconds to one minute regarding the seeing conditions. Recent external profiling experiments acquired the profile on-sky every 5\,mn \citep{Osborn2018} and alternative approaches emerges that aim to increasing the temporal resolution to 2\,mn \citep{Hickson2019}. Considering the fact that we need 30\,s-1\,mn-long observation with PEPITO depending on the detector configuration, we may have a factor 4 improvement on the temporal resolution.}.

We present the concept of PEPITO in Sect.~\ref{S:Concept} and detail the model of Point Spread Function (PSF) with respect to the $\cnh$ and $L_0(h)$. In Sect.~\ref{S:Simu}, we validate the approach and show that PEPITO is capable of retrieving the profile at better than 1\%-level of accuracy. Finally, we present in Sect.~\ref{S:sensitivity} a sensitivity analysis with respect to the PSF field location and noise.

\section{Atmospheric profile retrieval from seeing-limited PSF}
\label{S:Concept}	

\subsection{Digital anisokinetism}
\label{SS:PSF}	

The anisokinetism is introduced either digitally or from a tip-tilt compensator device. With the first solution, we recenter all the short-exposure images according to the measured position of a reference star and then average across them. The digital approach involves a simple optical design, i.e.~an imaging camera at the telescope focus but requires the collection of sufficient short-exposure frames to generate the long exposure equivalent and over a several arcminute FOV. The a processing pipeline must deal with $\sim10^4$ images, ${O}(10\,\mathrm{Gb})$, before the FPP algorithm can be applied. Using a tip-tilt compensator reduces the imaging camera's noise requirements and lower computational requirements but instead involves compensating the TT for a target in real time. An additional disadvantage of real-time compensation is only permitting one star to be the references within the field; the digital approach allows for any star in the field to be the reference if all short-exposure images are retained and this permits an increase in the number of anisokinetism realizations. In the following text we therefore rely on descriptions based on digitally created anisokinetism.\\

Therefore, we consider a data-cube of $n_\text{Exp}$ short-exposure frames containing at least two PSFs in the field. We assume that the image is contaminated by a zero-mean noise. To introduce the anisokinetism, we must calculate firstly the tip-tilt values according to the reference short-exposure PSF that we denote $\psf{0}^\text{SE}$. We have to isolate each PSF from each other, which is done by truncating the image thanks to a gate function $\Pi_n^\theta$ that conserves only the pixels within a squared box of size $n$ and centered around the angular position $\theta$, that is known from a catalog and for the observed asterism. We start the process by estimating the reference PSF flux $I_0$
\begin{equation} \label{E:flux}
I_0 = \sum_{k=1}^{n_\text{Exp}}\sum_{i,j}^{n} \Pi_n^0(x_i,y_j).\psf{0}^\text{SE}(x_i,y_j,k),
\end{equation}
where $x_i,y_j$ ranges from $-n/2$ to $n/2$  and refers to angular separation given in Cartesian coordinates that are defined around the reference PSF catalog position. We assume that there is no flux variability across the observation, which will be nominally kept shorter than a minute.

Then, we estimate the PSF barycenter $(\bar{x}_0(k),\bar{y}_0(k))$ frame-by-frame,
\begin{equation} \label{E:barycenter}
\begin{aligned}
&\bar{x}_0(k) = \dfrac{1}{I_0}\times\sum_{i,j}^{n} \Pi_n^0(x_i,y_j).\psf{0}^\text{SE}(x_i,y_j,k).x_i,\ \\
&\bar{y}_0(k) = \dfrac{1}{I_0}\times \sum_{i,j}^{n} \Pi_n^0(x_i,y_j).\psf{0}^\text{SE}(x_i,y_j,k).y_j.\\
\end{aligned}
\end{equation}
Finally, the PSF separated by an angular shift of $\theta$ from the reference is corrected for the on-axis TT and averaged over the temporal dimension as follows
\begin{equation} \label{E:PSFavg}
\begin{aligned}
\psf{\theta}(x_i,y_j) = \dfrac{1}{n_\text{Exp}}\sum_{k=1}^{n_\text{Exp}}& \Pi_n^\theta(x_i,y_j)\\&\psf{\theta}^\text{SE}(x_i-\bar{x}_0(k),y_j-\bar{y}_0(k),k).
\end{aligned}
\end{equation}
where $\psf{\theta}$ is the TT-corrected long-exposure PSF in the direction $\theta$ from the reference.

\subsection{PSF direct model}

\begin{figure*}
\centering
\includegraphics[width=18cm]{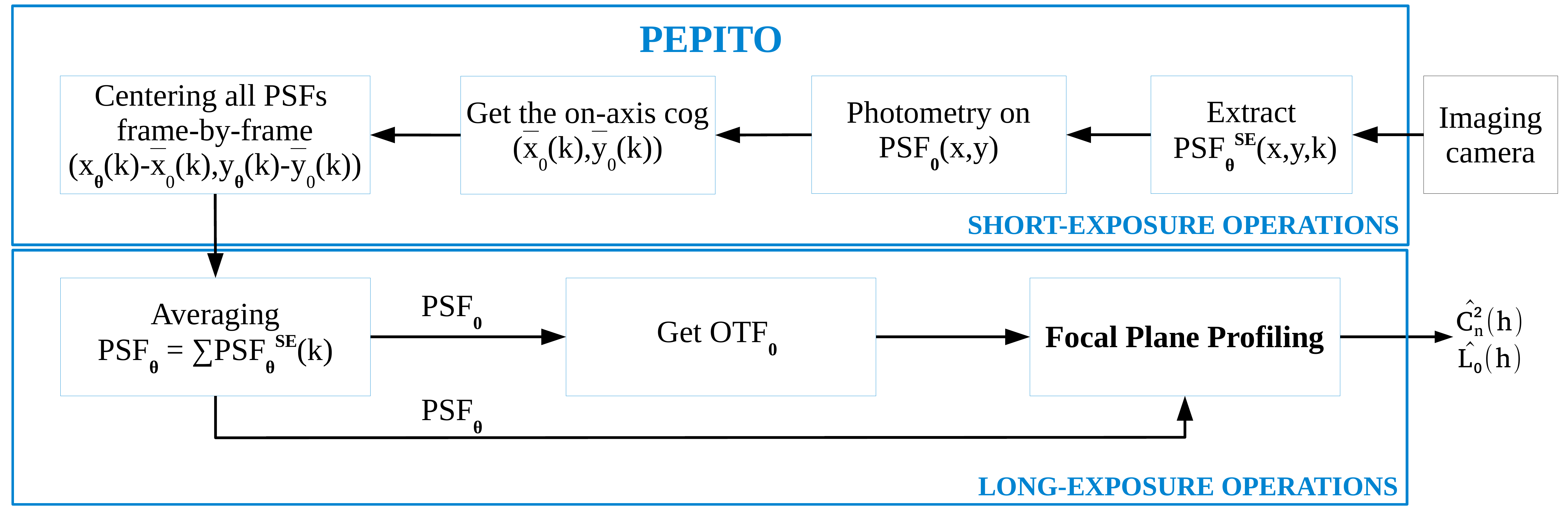}\caption{Schematic diagram of digital PEPITO. The algorithm starts by finding the barycentre (centre-of-gravity) for a reference target, PSF\textsubscript{0}, from short-exposure images, then recentering each image's extracted PSF based on the reference barycentre, and then forming a tip/tilt corrected long-exposure image per PSF from the centred short-exposure images. Off-axis PSF are consequently contaminated by anisokinetism that are provided to the FPP algorithm to retrieve the $\cnh$ and $L0$ profiles. }
\label{F:PEPITO}
\end{figure*}

Then, PEPITO derives the corresponding Optical Transfer Function (OTF) $\otf{\theta}$, from the following equation
\begin{equation}\label{E:model}
		\otf{\theta}(\rhob/\lambda,\cnh) =  \otf{0}(\rhob/\lambda) \cdot \atf{\theta}(\rhob/\lambda,\cnh),
\end{equation}
 where $\rhob$ is the separation vector of two phase samples within the pupil, $\otf{0}$ is the on-axis long-exposure OTF, and $\atf{\theta}$ is the Anisoplanatism Transfer Function introduced calculated from
\begin{equation} \label{E:atf}
\begin{aligned}
	\lefteqn{ \atf{\theta}(\rhob/\lambda,\cnh)  = } &\\
	& \dfrac{\iint_\pup \pup(\rbb)\pup(\rbb+\rhob)\exp	\para{-0.5\times \mathcal{D}_\Delta(\rbb,\rhob,\cnh,\theta)}\boldsymbol{dr}}{\iint_\pup \pup(\rbb)\pup(\rbb+\rhob)\boldsymbol{dr}}
		\end{aligned}
\end{equation}
where $rbb$ is the phase coordinates within the pupil,
$\mathcal{P}$ is the telescope pupil function and $\mathcal{D}_\Delta$ the anisokinetism structure function (SF). This latter characterises the TT decorrelation \citep{Conan2000} of two wavefronts coming from two stars separated in angle by $\theta$ and is derived as follows
\begin{equation} \label{E:Ddelta}
\begin{aligned}
\mathcal{D}_\Delta(\rbb,\rhob,\cnh,\theta)  &=  2\times\FTT.\left[ \mathcal{D}_0(\rbb,\rhob,\rz,L_0(h)) \right.\\
	&\left.-\mathcal{D}_\theta(\rbb,\rhob,\cnh,L_0(h),\theta)) \right].\FTT^t
	\end{aligned}
\end{equation}	
with $\mathcal{D}_0$ the atmospheric phase SF including all modes, $D_\theta$ the cross-correlated SF of the atmospheric phase for a separation of $\theta$ and $\FTT$ the matrix filter that conserves the TT modes only. $\mathcal{D}_0$ depends on integrated values of $\cnh$ and $L_0(h)$ profiles over altitude, not on the specific height distribution. The calculation of $\mathcal{D}_\Delta$ is performed thanks to the anisoplanatism model \citep{BeltramoMartin2018_aniso} included into the OOMAO simulation framework \citep{Conan2014OOMAO}. The algorithm architecture is summarized in Fig.~\ref{F:PEPITO}

\section{Simulation-based validation}
\label{S:Simu}	

\subsection{Simulations of digital anisokinetism}
\label{SS:simu}		
We have simulated 50000 short-exposure V-band images of 10\,ms exposure within a 8 arcmin FOV using the end-to-end simulator OOMAO, with a 0.5\,m and 1\,m circular telescope with 30\% central obscuration. The atmosphere is simulated over 7 layers at 0.1, 0.5, 1, 2, 4, 8 and 16 km with strengths of respectively 60.4, 11.3, 9.6, 2.9, 5.8, 4.2, \& 5.8\% and a Fried's parameter of $r_0=15.7$\,cm, as it represents the median seeing conditions at MaunaKea . The zenith angle is 30${}^\circ$ and the outer scale was fixed to $L_0=25$\,m. \textcolor{black}{Also, simulations were relying on a frozen-flow turbulence with a 5000Hz temporal frequency to sample adequately the temporal stochastic variations of the atmospheric phase up to wind speed of 50\,m/s.}

\begin{figure}
\centering
\includegraphics[width=8.25cm]{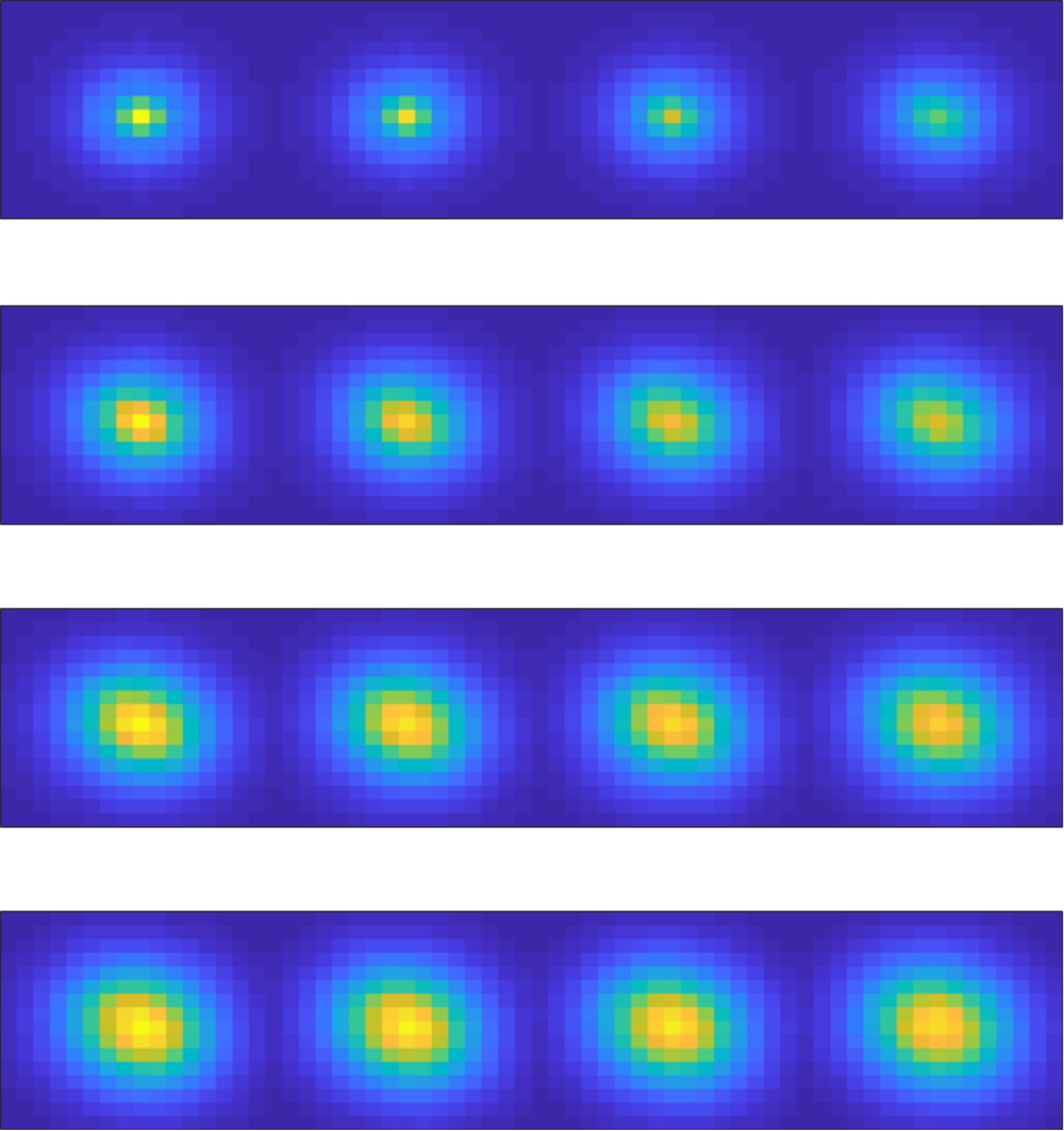}
\caption{Simulated on-axis TT-corrected PSF for $\theta=0$ (top-left) to $30"$(bottom-right), averaged out over 50s. The scale changes to reveal the elongation variations more clearly. }
\label{F:psfVtheta}
\end{figure}

\begin{figure}
\centering
\includegraphics[width=9.5cm]{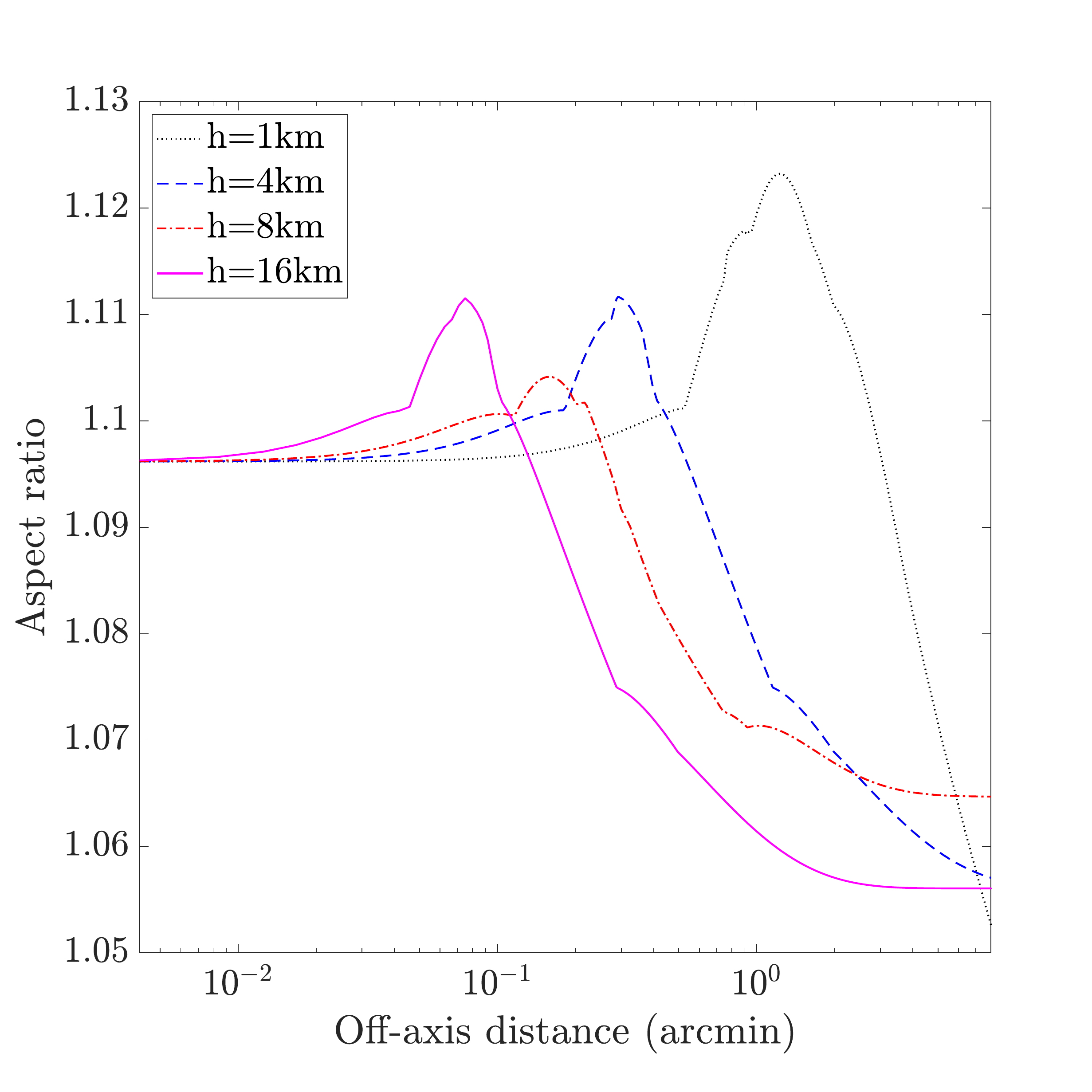}
\includegraphics[width=9.5cm]{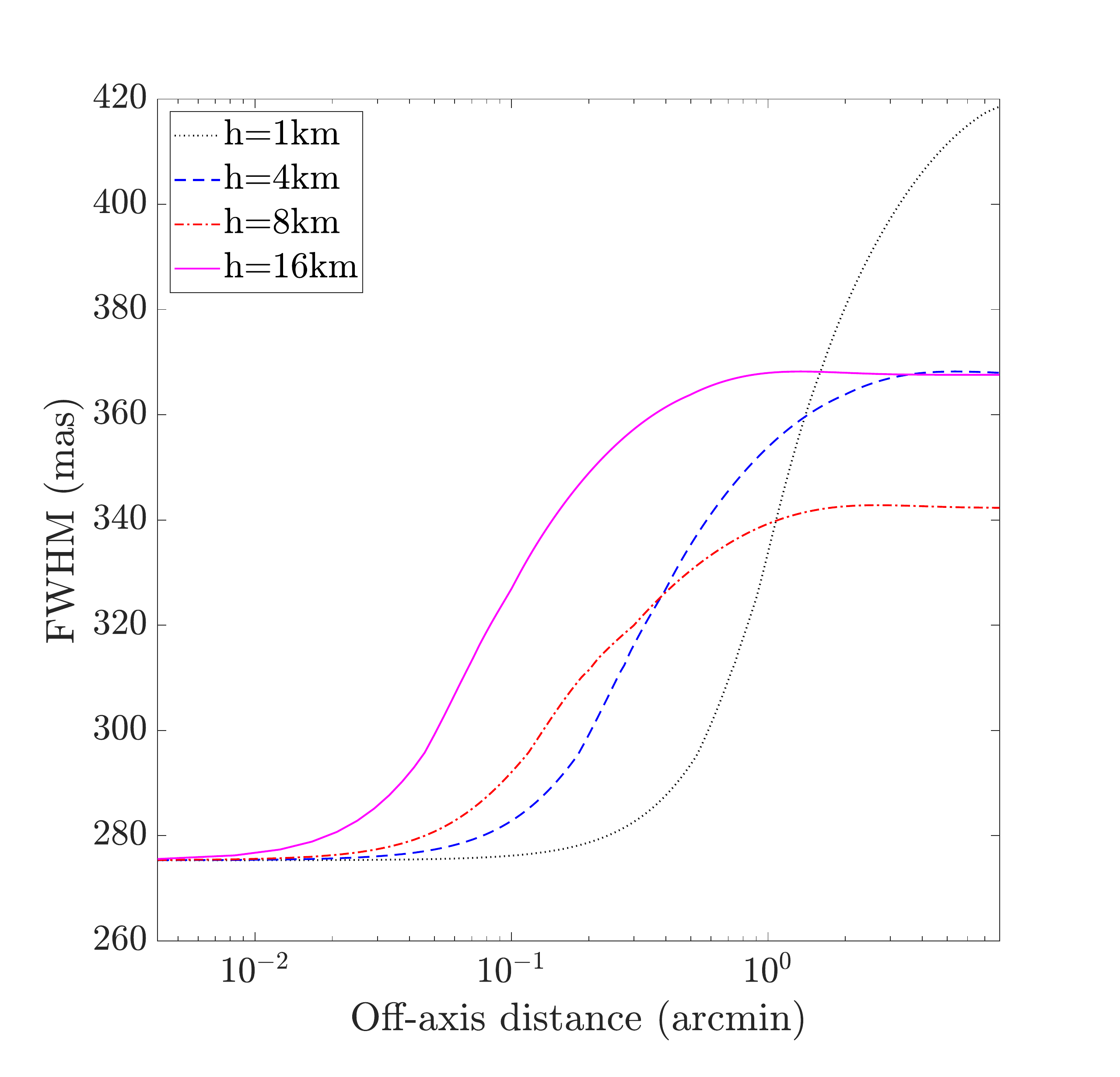}
\caption{{\em Top:} PSF aspect ratio, and {\em bottom:} \textbf{FWHM} for anisokinetism from a single atmospheric turbulence layer at one of four test altitudes. Varying layer strengths cause saturation and maximal elongation to different values.}
\label{F:aspectRatioVdist}
\end{figure}

Fig.~\ref{F:psfVtheta} illustrates the simulated PSFs with respect to their position in the field from 0\," to 30\,", showing that the anisokinetism, and consequently the atmospheric profiles, affect the PSF elongation and FWHM. In Fig.~\ref{F:aspectRatioVdist}, we investigate those metrics as function of $\theta$ for a single layer at different heights. The plots shows that the PSF elongation reaches a maximum value at a value of $\theta$ which depends on the layer height, while the FWHM grows up monotonically.

For small separations, anisokinetism does not significantly affect the PSF morphology. On the contrary, for larger separations we observe a saturation of the effect when the FWHM reaches a plateau that comes with a diminution of the aspect ratio. This ratio eventually drops to one indicating that the PSF becomes centro-symmetric. We also notice that the maximum aspect ratio coincides with the center of the FWHM linearity zone in Fig.~\ref{F:aspectRatioVdist}. 

Furthermore, the anisokinetism signature of higher altitude layers occurs for shorter separations: PEPITO will retrieve higher altitude layers within a smaller field-of-view. This FOV is instead constrained by the {\em minimum} altitude we want to characterize. On top of that, the FWHM can only increase with respect to $\theta$ because the anisokinetism saturation we see on PSF FWHM in Fig.~\ref{F:aspectRatioVdist}. Therefore, the estimation of $\cnh$ in lower altitude layers will rely on broader PSFs at larger separations than estimation for higher altitudes whose anisokinetism will have saturated the FWHM. Thus a full profile estimation based on multi-PSF fitting is necessary.

\subsection{Model verification}

We propose to validate Eq.~\ref{E:model} which estimates the OTF, and consequently the PSF model, with respect to $\theta$ and the $\cnh$ profile. According to the simulation parameters, we have computed the ATF and the corresponding PSF for all values of $\theta$ from 0\," to 30\," as they are displayed in Fig.~\ref{F:psfVtheta}. The PSF computation is fast and less than 1\,s-long on a regular laptop.

\begin{figure}
\centering
\includegraphics[width=9.5cm]{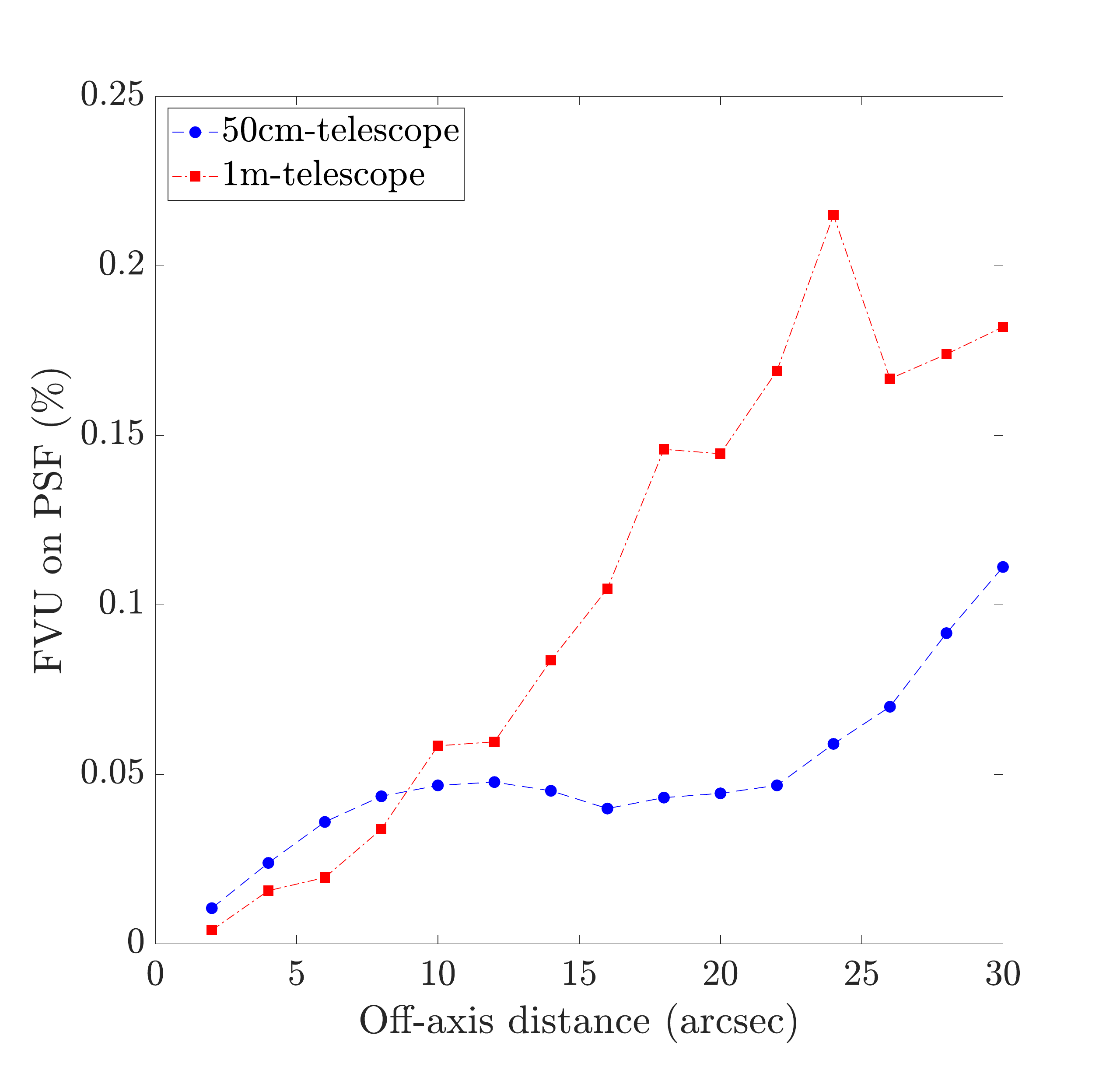}
\caption{Fraction of variance unexplained on the PSF model compared to simulations with respect to the PSF field position and for a 0.5\,m- and 1\,m-class telescope.  }
\label{F:fvuPSF}
\end{figure}

\begin{figure*}
\centering
\includegraphics[width=18cm]{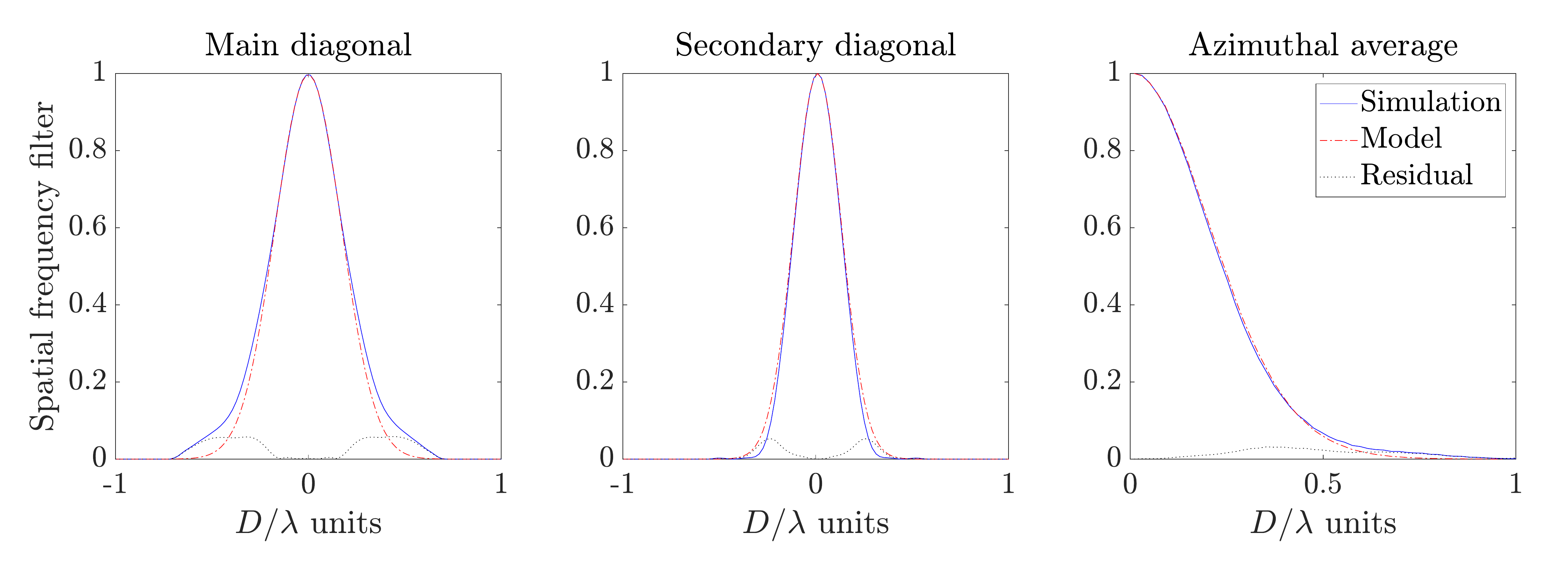}
\includegraphics[width=18cm]{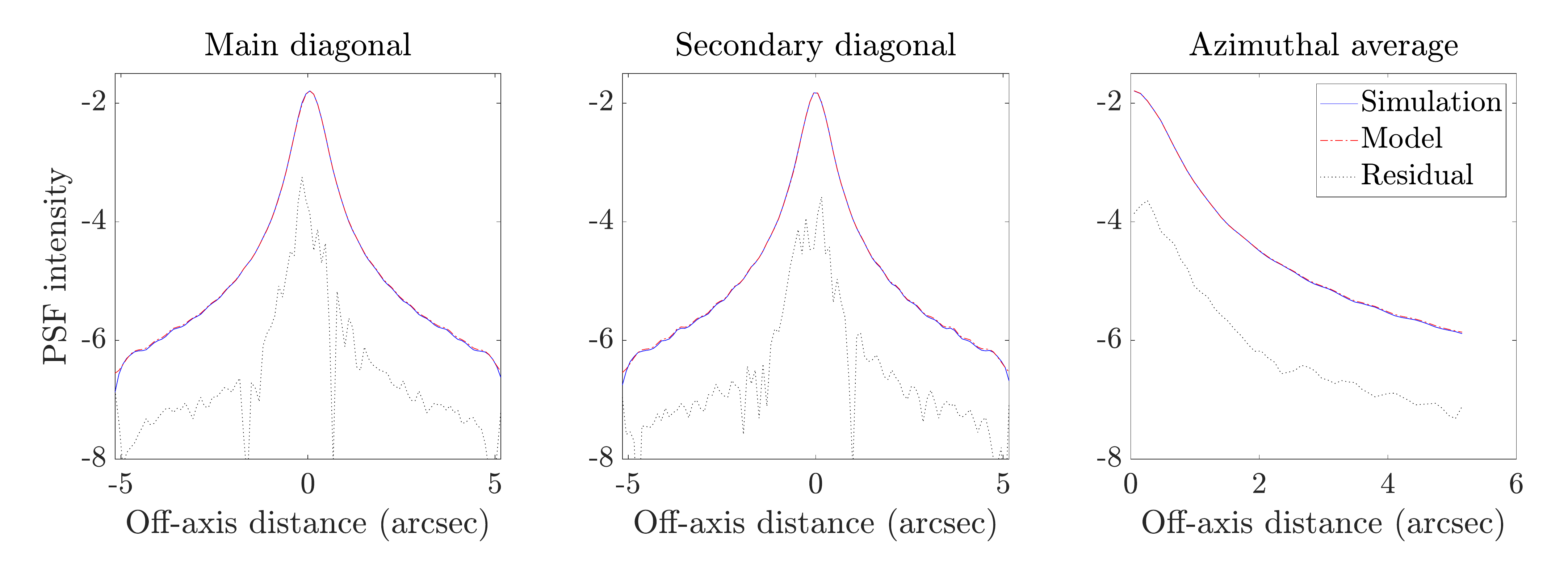}
\caption{\textbf{Top:} ATF and \textbf{bottom:} PSF 1D profile delivered by the simulation for $\theta=16"$ and compared to the model. The main diagonal is obtained in the maximal elongation direction and the secondary diagonal is the one perpendicular to.}
\label{F:psf1Dplots}
\end{figure*}

In Fig.~\ref{F:psf1Dplots} are shown the ATF and PSF for $\theta=16"$, which highlight an excellent agreement. We have verified that we get the same level of accuracy regardless of $\theta$ and with the same accuracy on the PSF elongation. We report in Fig.~\ref{F:fvuPSF} the Fraction of Variance Unexplained (FVU) that measures the overall model error over all angular separations \citep{BeltramoMartin2018_aniso}. The figure emphasizes that the residual model error stay below 0.2\% for two different values of telescope diameter and over a 30"-field of view. For larger separations, the PSF becomes more and more centro-symmetric because the saturation effect introduced by low altitude layers, as we see it in Fig.~\ref{F:psfVtheta}, whose the shape is well predictable and justifies why we concentrate our analysis on this particular 30" range.

As a conclusion, the anisokinetism model proposed in Eq.~\ref{E:Ddelta} is a sufficiently accurate description to mimic the impact of the on-axis TT-correction on long-exposure PSF distributed over the field. Consequently, PEPITO can use this model and exploit the TT-corrected PSF in the field to invert the problem and retrieve the atmospheric profiles.
 
\subsection{Retrieval performance}
\label{SS:perf}

\begin{figure}
\centering
\includegraphics[width=9.5cm]{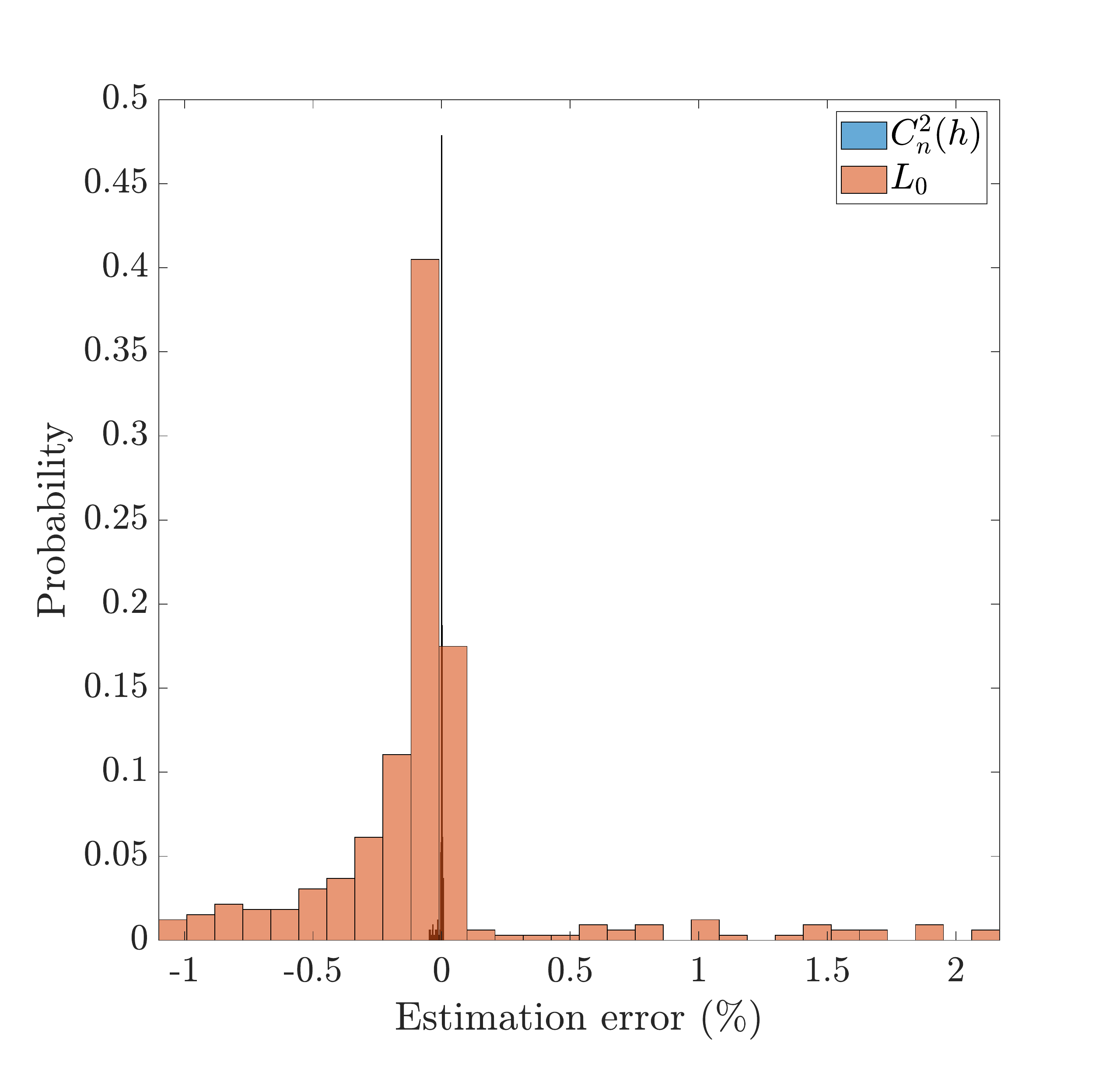}
\caption{\small Error probability on the $\cnh$ and $L_0$ given for a various random initial guess. The $\cnh$ and $L_0$ distributions has respectively -0.001\,\%/-0.05\,\% of mean error and 0.01\,\%/0.5\,\% of 1-$\sigma$ standard-deviation.}
\label{F:histoCn2herrorVinit}
\end{figure}

We have selected 7 PSFs at particular positions (see Sect.~\ref{SS:SensitivityPos}) and pass them to the FPP algorithm to retrieve the $\cnh$ and $L_0$ profiles, with random initial guesses. Systematically, an atmospheric profile estimation is obtained with an accuracy at the tenth of a percent level on the $\cnh$ and one percent on the $L_0$, as presented in Fig.~\ref{F:histoCn2herrorVinit}. \textcolor{black}{The $\cnh$ relative error distribution is strongly peaked at zero with a mean value of -0.001\%, with 0.01\,\% of standard-deviation, and a probability of one to remain below 0.3\,\%. The relative precision of $L_0$ retrieval has a mean deviation of -0.05\,\% and 1-$\sigma$ uncertainty of 0.5\,\%, with a probability of 0.93 to stay below 1\% of error. We have verified that these precisions for a range of of $\cnh$ and $L_0(h)$ values. Therefore we conclude that $\cnh$ and $L_0$ can be retrieved with relative accuracy of 0.3\% and 1\% from anisokinetism-affected PSFs.}

\begin{figure}
\centering
\includegraphics[width=9.5cm]{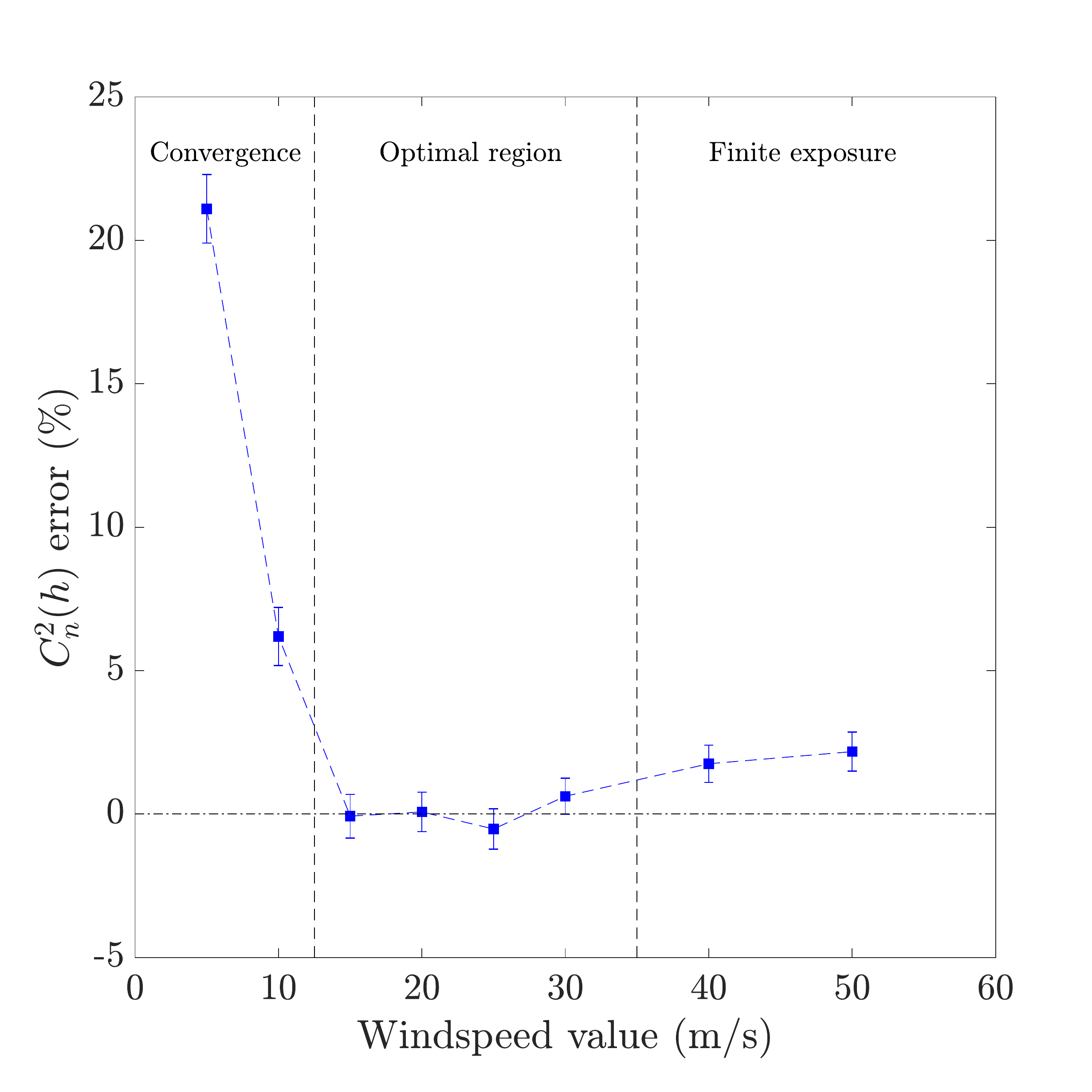}
\caption{\small  \textcolor{black}{$\cnh$ error as function of the wind speed in simulations. The plot highlights three different regimes: error dominated by the lack convergence of atmosphere statistics, optimal regime and impact of the finite exposure time.}}
\label{F:cn2hErrorVwindspeed}
\end{figure}

\textcolor{black}{On top of that, we have also analyzed how the $\cnh$ accuracy varies with respect to the wind speed value, by considering a sole layer at 16\,km, as reported in Fig.~\ref{F:cn2hErrorVwindspeed}.  The estimates error reaches the value presented in Fig.~\ref{F:histoCn2herrorVinit} for a regime of wind speed from 15\,m/s to 30m/s. For a lowest turbulence, the $\cnh$ error increases due to a lack of atmosphere statistics convergence, as it will be more discussed in Sect.~\ref{SS:tempResolution}. This can be solved by integrating longer. For higher wind speed values than 30m/s, the error slightly goes up to 2\,\%, but for a different reason: because the finite exposure time of 10\,ms, the tip-tilt estimation becomes less accurate when the atmosphere coherence time get shorter. The time averaging serves as a low-pass filter~\citep{Martin1987,Hickson2019} that blurs the PSF and introduces an additional component that superimposes to the TT-anisoplanatism effect and diminishes the PEPITO sensitivity.}

\textcolor{black}{However, according to~\citep{Hickson2019}, the important scalar parameter to focus on is the atmosphere characteristic time defined as $T_0 = \pi.D/v_{8/3}$, where:
\begin{equation}
    v_{8/3} = \para{\dfrac{\int_0^\infty \cnh v(h)^{8/3}dh}{\int_0^\infty \cnh dh}}^{3/8}.
\end{equation}
For a 1m-telescope and with $T=10$\,ms as the exposure time, we get $T/T_0$ = 0.1 for respectively a wind speed value of 30\,m/s, which correspond to the limit presented by~\citep{Hickson2019} to consider the finite exposure time as negligible in the seeing estimation. It coincides with our present results showing the estimates accuracy degradation up to 2\,\% for faster wind speed than 30\,m/s. According to recent surveying of the atmospheric profile at Paranal~\citep{Osborn2018,Masciadri2014}, having a layer that combines high speed ($>$ 30\,m/s) and large strength ($>$5\,\% of the whole profile) as well as high altitude ($>$10\,km) to produce a detectable anisoplanatism signature is not frequent, advocating for a mitigation of this effect with PEPITO, especially regarding the convergence issue that is the main constrain that will specify the exposure time.}

\section{Sensitivity analysis}
\label{S:sensitivity}
\subsection{Temporal resolution}
\label{SS:tempResolution}

\begin{figure}
\centering
\includegraphics[width=9.5cm]{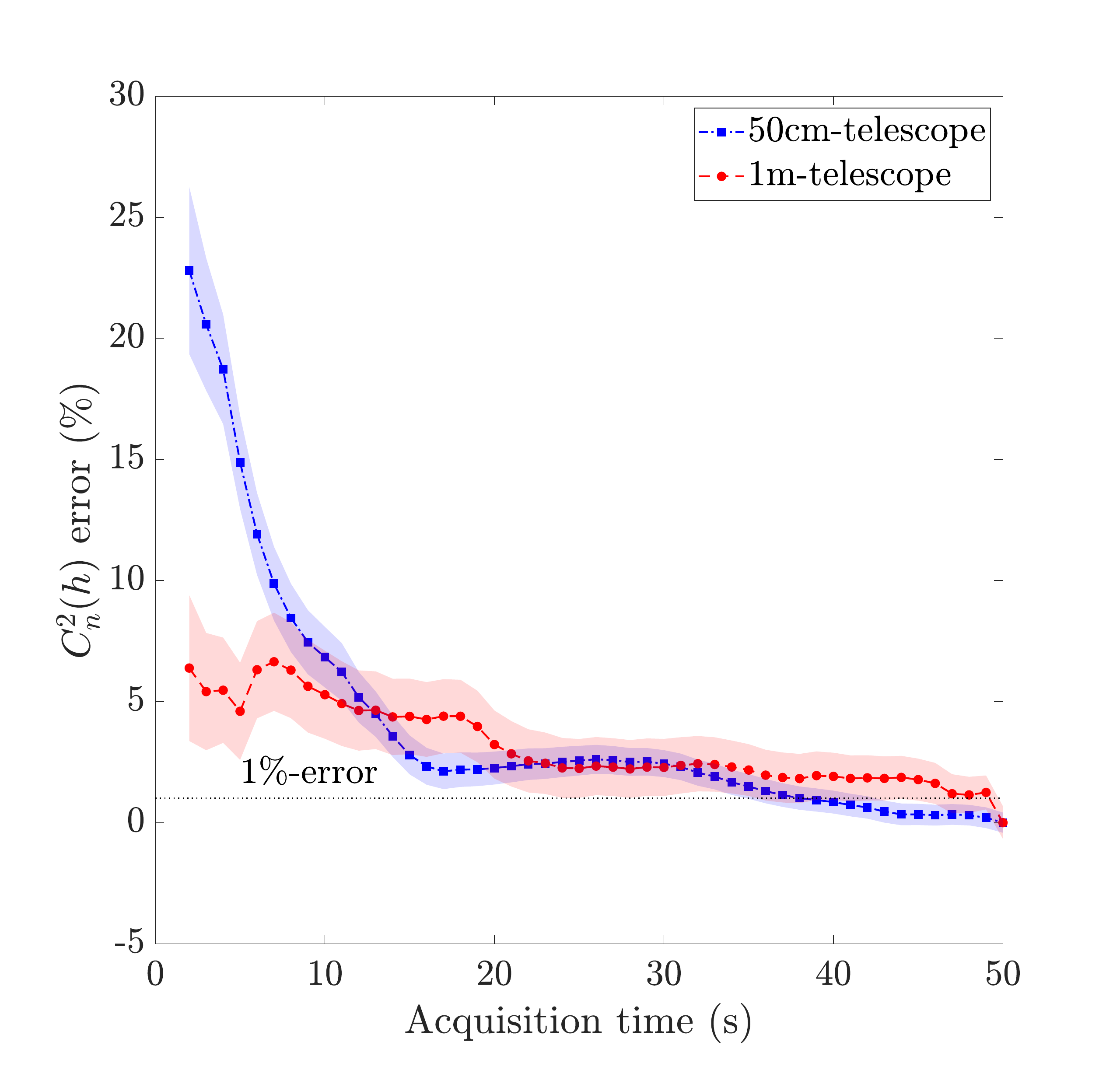}
\caption{Relative error on the $\cnh$ estimation with respect to the acquisition time. Envelopes give the 1-$\sigma$ precision.}
\label{F:cn2hErrorVtime}
\end{figure}

The temporal resolution is limited by the exposure time required for the long exposure PSF to converge to its average \citep{Gordon2011}. 
From the noise-free PSFs simulated with 0.5\,m and 1\,m diameter telescopes, we have investigated the error in estimating the $\cnh$ profile versus integration time between 2\,s and 50\,s, as reported in Fig.~\ref{F:cn2hErrorVtime}. The figure demonstrates that to reach an accuracy of 1\% requires at least 30\,s exposure time, with less influence from the telescope diameter compared to conclusions presented in\citep{Gordon2011}. We see several reasons to this, such the TT correction that removes the strongest part of the atmospheric distortions and the PSF fitting process that exploits efficiently pixels intensity to estimate the $\cnh$.

In the presence of noise, it is necessary to increase the exposure time to meet the signal-to-noise requirements presented in Sect.~\ref{SS:SensitivityNoise}. This value of 30s exposure is the minimal duration to reach a convergence of the atmosphere statistics that produces meaningful estimations in the visible. In other words PEPITO's temporal resolution is limited by the PSF convergence to its long exposure expectation.

\subsection{Field of view}
\label{SS:SensitivityPos}

We describe in the following a sensitivity analysis to understand the performance of PEPITO and highlight which conditions must be met to ensure an accurate atmospheric profile estimation.

\begin{figure}
\centering
\includegraphics[width=9.5cm]{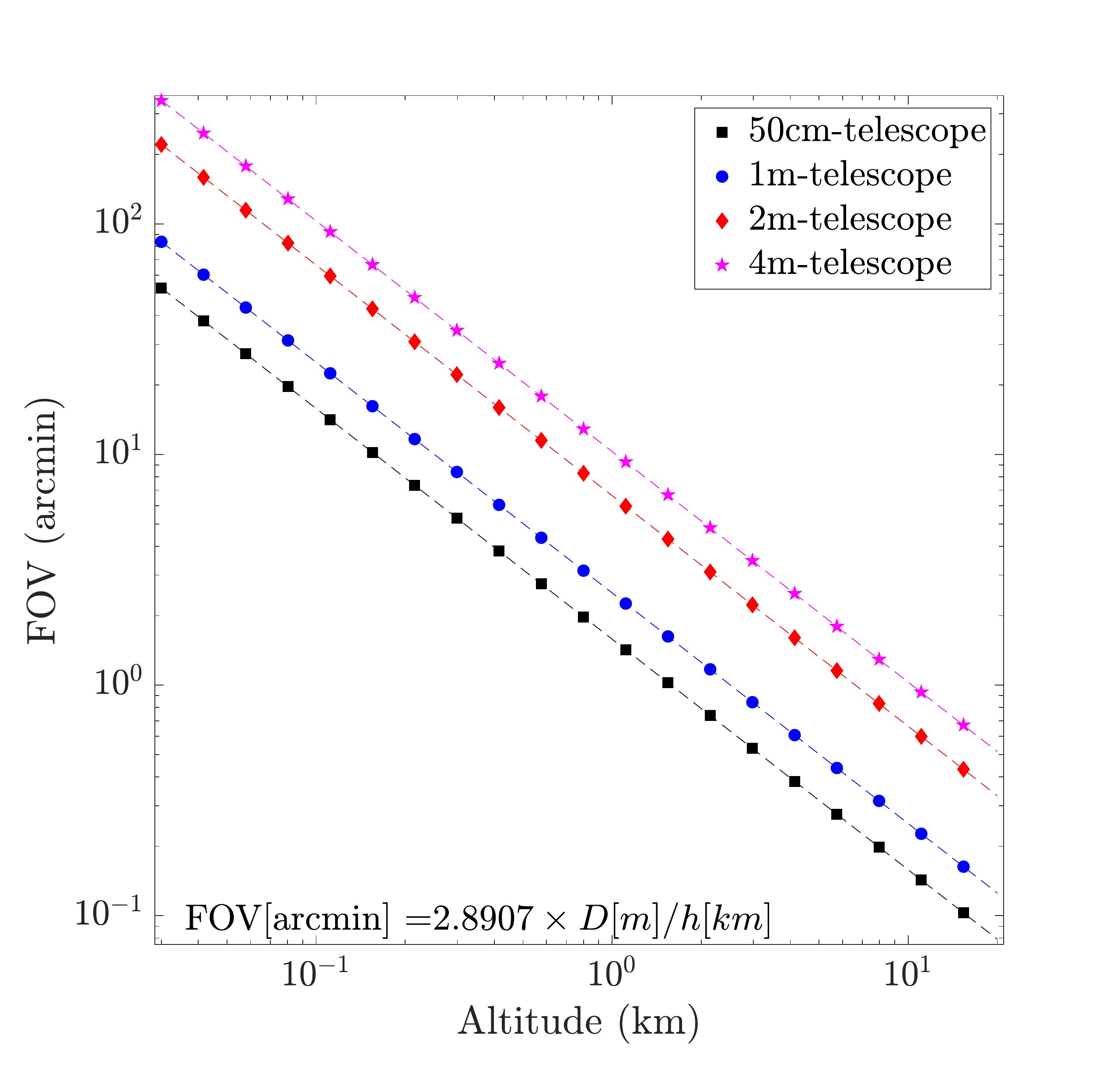}
\caption{\small Required field of view with respect to the layer height to be retrieved for a different telescope diameter. Markers are obtained from the sensitivity analysis and dashed lines correspond to linear regression.}
\label{F:locValt}
\end{figure}

According to Eqs.~\ref{E:atf} and~\ref{E:Ddelta}, the relation between the PSF morphology, the atmosphere distribution and the telescope configuration is highly non-linear and quite complex. Therefore we have assessed empirically the PSF location for which the maximum aspect ratio is found as function of the layer height and telescope diameter, and presented in Fig.~\ref{F:locValt}. We have simulated a single layer profile and spanned the altitude range in order to measure the required FOV to capture the largest signature of the anisokinetism, i.e.~where the PSF aspect ratio is maximal. We ended up with the following expression
\begin{equation} \label{E:thetaopt}
\text{FOV}(h) \cro{\text{arcmin}} = \alpha\times\dfrac{D\cro{\text{m}}}{h\cro{\text{km}}},
\end{equation}
where $\theta_\text{opt}$ is the optimal PSF position in arcmin for the layer height $h$, in km, and with $\alpha = 2.9 \pm 0.2$ as an empirical factor estimated from a data fitting. We can also evaluate the minimum altitude that PEPITO can resolve as function of the available FOV from 
\begin{equation}
h_\text{min} = \alpha\times\dfrac{D}{\text{FOV}},
\end{equation}
which gives $h_\text{min} =150\,$ m for a 10' field of view on a 0.5\,m telescope.

\subsection{Altitude resolution}
\begin{figure}
\centering
\includegraphics[width=9.5cm]{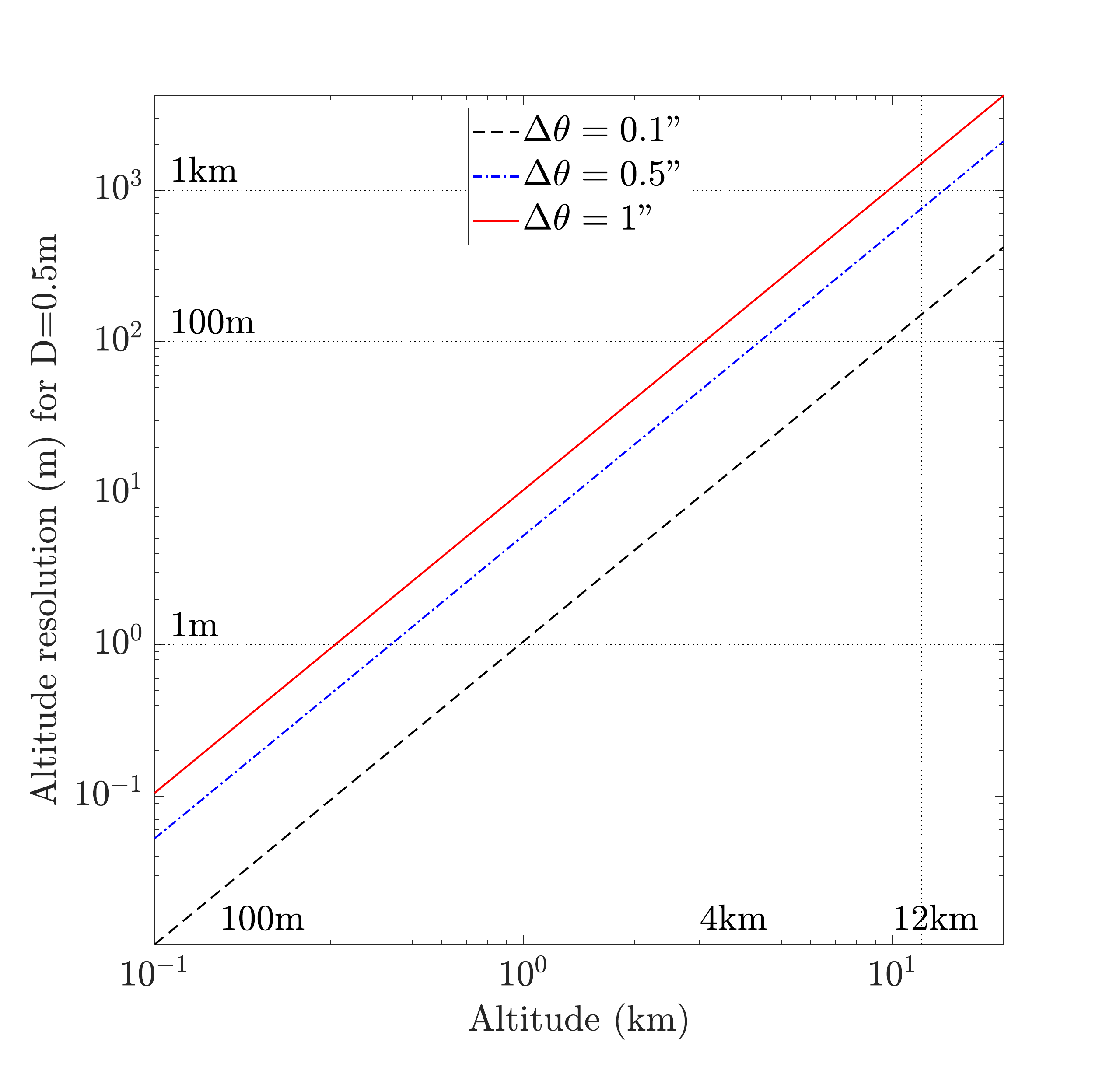}
\caption{PEPITO altitude resolution with respect to the layer height for various pixel scale values $\Delta \theta$.}
\label{F:altitudeRes}
\end{figure}

From the derivative of Eq.~\ref{E:thetaopt} with respect to $h$, we show that the altitude resolution increases as a function of $h^2$ as follows
\begin{equation}
\Delta h = \dfrac{\Delta\theta}{\alpha{}D} \times h^{2}
\end{equation}
where $\Delta\theta$ is the detector pixel scale in arcmin, which gives the shortest angular separation measured on the detector. The altitude resolution therefore degrades quadratically with respect to the layer height, as illustrated in Fig.~\ref{F:altitudeRes}.

The SCIDAR technique has an altitude resolution $\delta h$ that varies with respect to $\sqrt{h}$ \citep{Farley2018}, but is limited by the separation of the two stars it relies on. Instead, PEPITO allows for an adjustment of altitude resolution from 1\,m to 400\,m by changing the camera pixel scale.  Although PEPITO aims particularly to calibrating the anisoplanatism model for SCAO-assisted large-field observations depleted into sufficiently bright point sources, there is a potential interest to use it on multiple guide stars systems as well. Thanks to the adjustable resolution, we could deploy PEPITO to enhance the high altitude layers characterization and optimize the system control regarding both the telemetry and the focal-plane image as well. Such a possibility will be investigated in the future.

\subsection{Layer completeness}
\label{SS:completeness}

In \citep{BeltramoMartin2018_FPP}, we have shown that the $\cnh$ characterization is possible by a single AO-compensated PSF. The anisoplanatism effect manifests itself on a large number of modes which makes all the focal plane pixels meaningful for the profile estimation. In the present case of PEPITO, the anisokinetism only affects the PSF with the tip mode and tilt mode; each PSF provides a single measurement within the profile which suggests that the number of PSFs required equals the number of layers to be retrieved.

\begin{figure}
\centering
\includegraphics[width=9.5cm]{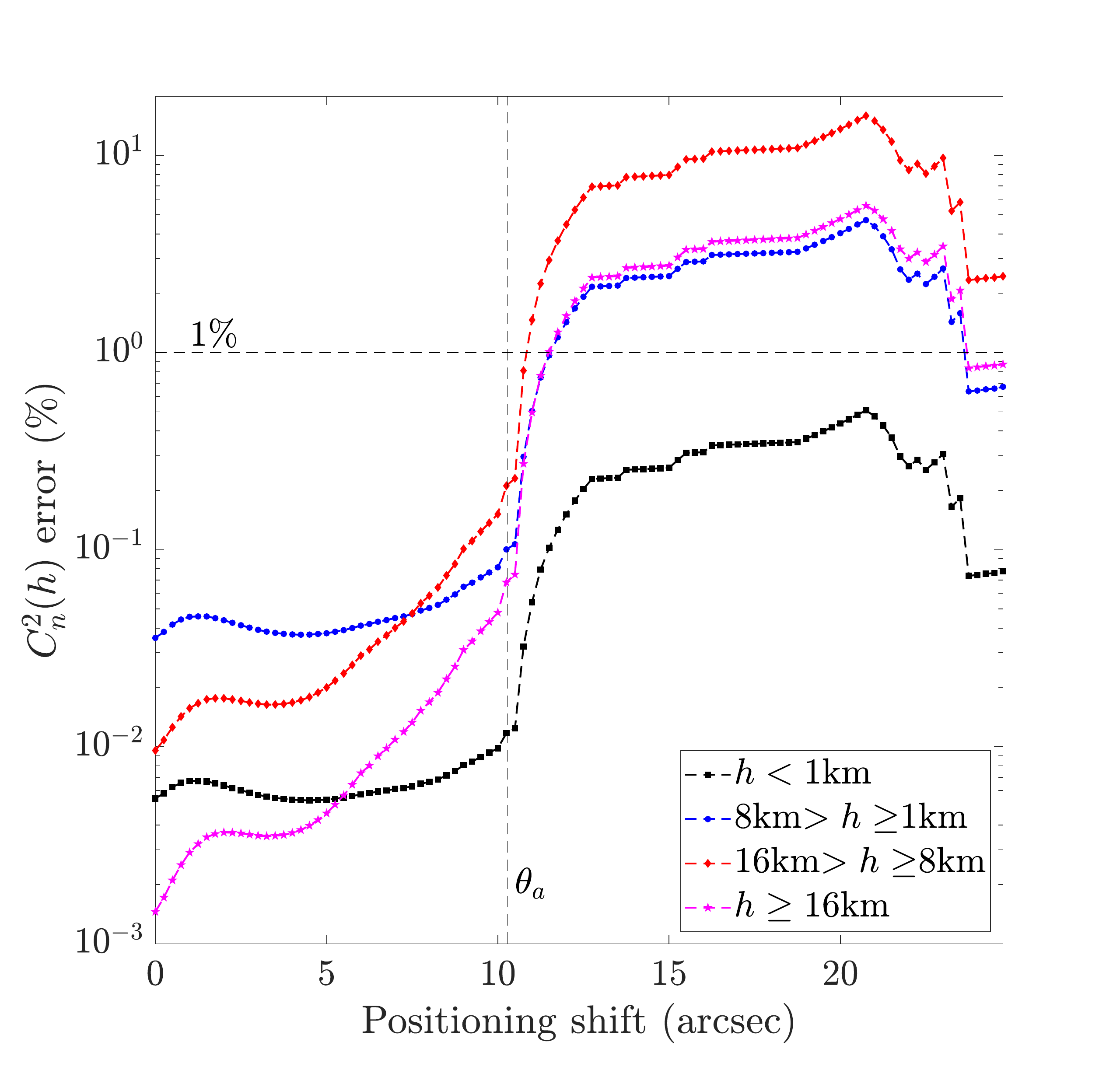}
\caption{Error on the $\cnh$ estimation regarding the PSF position shift from the optimal location.}
\label{F:cn2hErrorVshift}
\end{figure}

A related question is what the width of the angular region is in order to be capable of identifying a layer at altitude $h$? We made the exercise to pass to PEPITO a PSF shifted from its optimal position, and then measured the estimation error, as reported in Fig.~\ref{F:cn2hErrorVshift}. The plot gives evidence that there is a particular angle, $\theta_\text{a}$, beyond which the estimation integrity is compromised. The value of $\theta_\text{a}$ corresponds to the anisokinetism angle \citep{Roddier1981} which is the angular separation for wavefront variance from anisokinetism alone is 1\,rad$^2$. This indicates us that a layer of height $h$ can be retrieved with a PSF located at $\alpha\times D/h \pm \theta_a$. Layer taken individually should be accessible within an angular range that depends on the layer height; however an estimation error on a particular layer will affect the retrieval of the entire profile. Defining this range according to $\theta_a$ ensures to achieve a correct simultaneous estimation over all layers.  \\
In other words, a PSF separation of $\theta$ in the focal plane permits the sampling of the vertical profile within the optimal altitude range $h = \alpha\times D/\theta$ and with a corresponding half-width given by
\begin{equation}
\label{E:width}
    w_h = \alpha\times \dfrac{D}{2\times\theta_a}
\end{equation}

According to Eq.~\ref{E:thetaopt}, we have a direct relation between the PSF optimal field position  and the altitude altitude. This equation and the last analysis raises a question of disentangling contributions from two layers: in the case where two layers are maximally affecting the PSF in a region tighter than $\theta_a$, can we retrieve both these layers ?

\begin{figure}
\centering
\includegraphics[width=9.5cm]{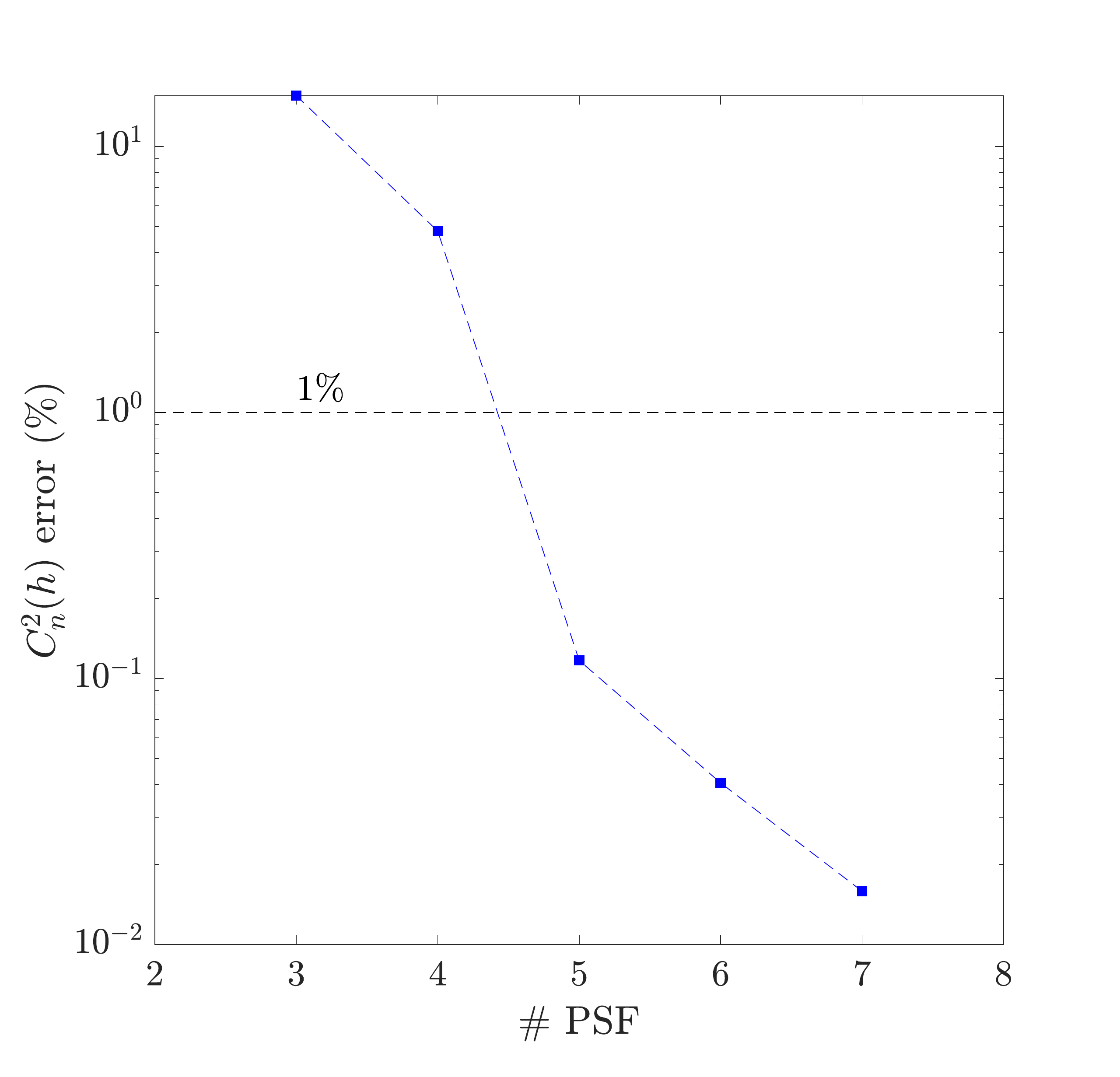}
\caption{Error on the $\cnh$ estimation with respect to the number of extracted PSFs. }
\label{F:cn2hErrorVnpsf}
\end{figure}

The simulated profile described in Sect.~\ref{SS:simu} leads to this situation. When deriving the optimal PSF position from the layers height, we get FOV$ = 323", 142", 74", 28", 18", 10", \& 5"$ for respectively $h = 0.1, 0.5, 1, 2, 4, 8, 16$ km. Regarding that $\theta_a = 10"$, this claims that we only need 5 PSFs to identify the full profile. We made the exercise of reducing the number of PSFs passed to PEPITO by ensuring that the PSFs locations were covering the optimal positions listed above within a range of $\pm \theta_a$. Fig.~\ref{F:cn2hErrorVnpsf} illustrates that PEPITO ensures a $\cnh$ estimation at 0.1\% of accuracy by using 5 PSFs, located at 323", 142", 74", 23", \& 7.5". When trying to reduce again the number of PSF, by substituting the two PSFs close to the reference by a single one positioned at 15", we found out that the problem is not sufficiently constrained to allow PEPITO to disentangling each layer.

\begin{figure}
\centering
\includegraphics[width=9cm]{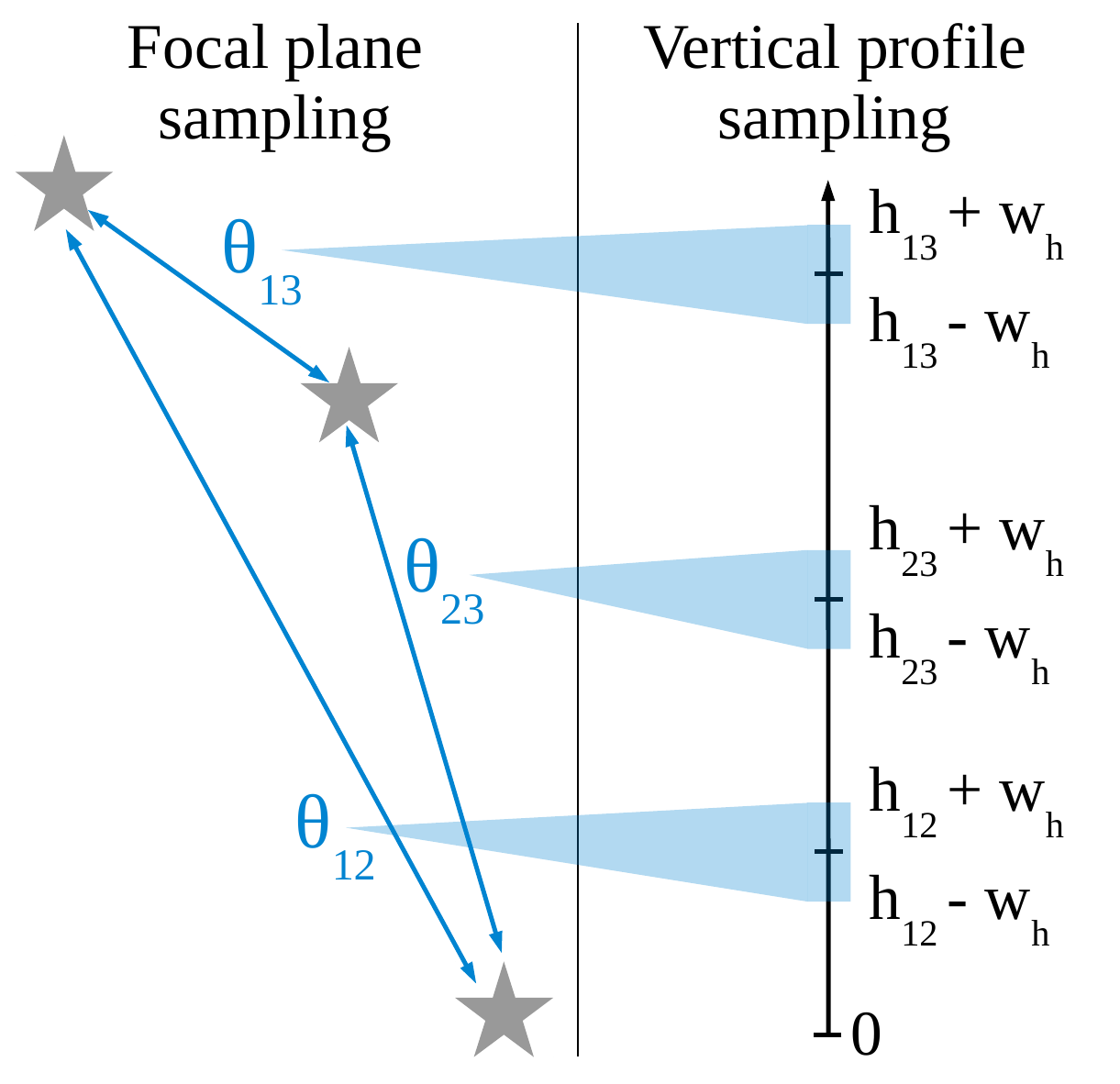}
\caption{Illustration of the possible baselines by choosing each star in turn as the reference. Each baseline permits the estimation of the atmospheric profile at specific height range whose the width $w_h$ in given by Eq.~\ref{E:width}.  In practice, each baseline has a second, redundant but independent measurement when the reference star and the anisokinetically affected star switch place e.g.~$h_{12}\equiv{}h_{21}$.
}
\label{F:baseline}
\end{figure}

We can therefore increase the signal-to-noise ratio by compensating the TT according to each PSF we have in the field to create different realizations of digital anisokinetism. Fig.~\ref{F:baseline} illustrates that each baseline formed by non-redundant PSFs pairs allows to cover a specific angular region that gives access to any layers comprised within the altitude range $w_h$. From a collection of $n_\text{PSF}$ PSFs distributed randomly in the field, assuming that all PSFs have baselines which are non-redundant, we can consequently identify $n_L$ layers where
\begin{equation}
n_L \geq \dfrac{n_\text{PSF}\times(n_\text{PSF}-1)}{2},
\end{equation}
that results in $n_L=15$ and $n_L=45$ layers respectively from $n_\text{PSF}=6$ and $n_\text{PSF}=10$ PSFs.

Having redundant baselines mitigates the noise contamination as well as reduces effect of static aberrations. Because we may need a large field of view to quantify low altitude layers, we can face spatial variation of static aberrations \citep{Sitarski2014}. In other words, term $\otf{0}$ in Eq.~\ref{E:model} contains a pattern due to the static aberrations in the on-axis direction, that changes when choosing another reference PSF and impact all the long-exposure PSF in the field. Consequently, a same baseline may provide a $\cnh$ estimation sightly different regarding the star we choose as a TT reference. Therefore accumulating redundant measurements is fundamental to reduce as much as possible the influence of static aberrations. It would also be helpful to refine the PSF model by calibrating the field static aberrations using a phase retrieval technique \citep{Lamb2016,Mugnier2008}.

\subsection{Sensitivity to noise}
\label{SS:SensitivityNoise}

Our last analysis relates to the noise contamination that has been overlooked so far. We consider a single layer at 10\,km and simulate the PSF at the field position given by Eq.\ref{E:thetaopt}. The PSF is scaled regarding its magnitude with a zero point set to $995\times10^6$ ph/m$^2$/s. We then include Poisson noise, sky background (mag V=24), read-out noise ($\sigma_e=0.8\text{e}^{-1}$) and dark current (0.05$\text{e}^{-1}$/s). We choose a detector quantum efficiency to be 70\% and an overall throughput of 90\% (prime focus). Finally, the long exposure-equivalent integration time was fixed to 30\,s by using 3000 short-exposure frames of 10\, ms duration. 

\begin{figure}
\centering
\includegraphics[width=9.5cm]{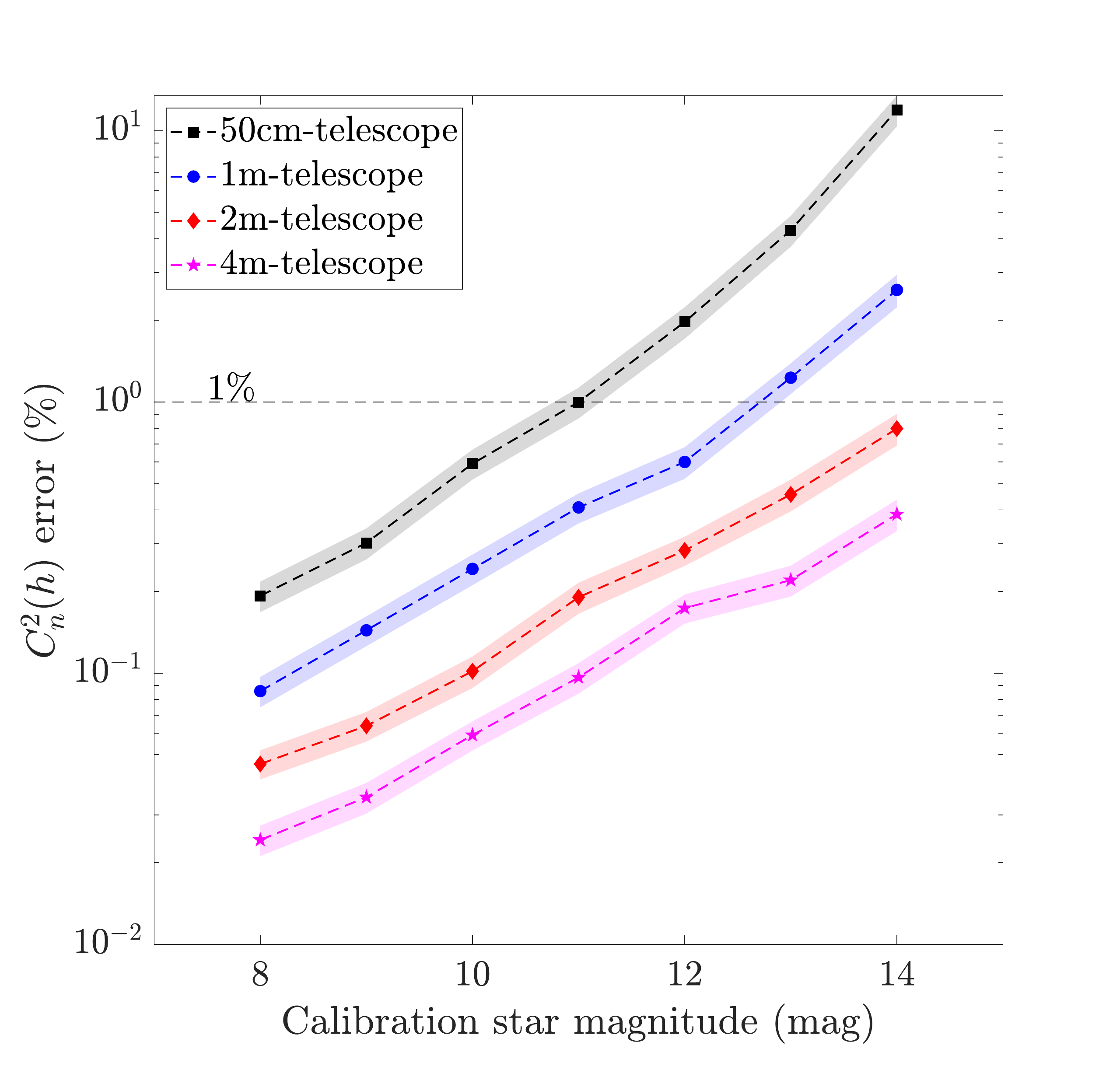}
\includegraphics[width=9.5cm]{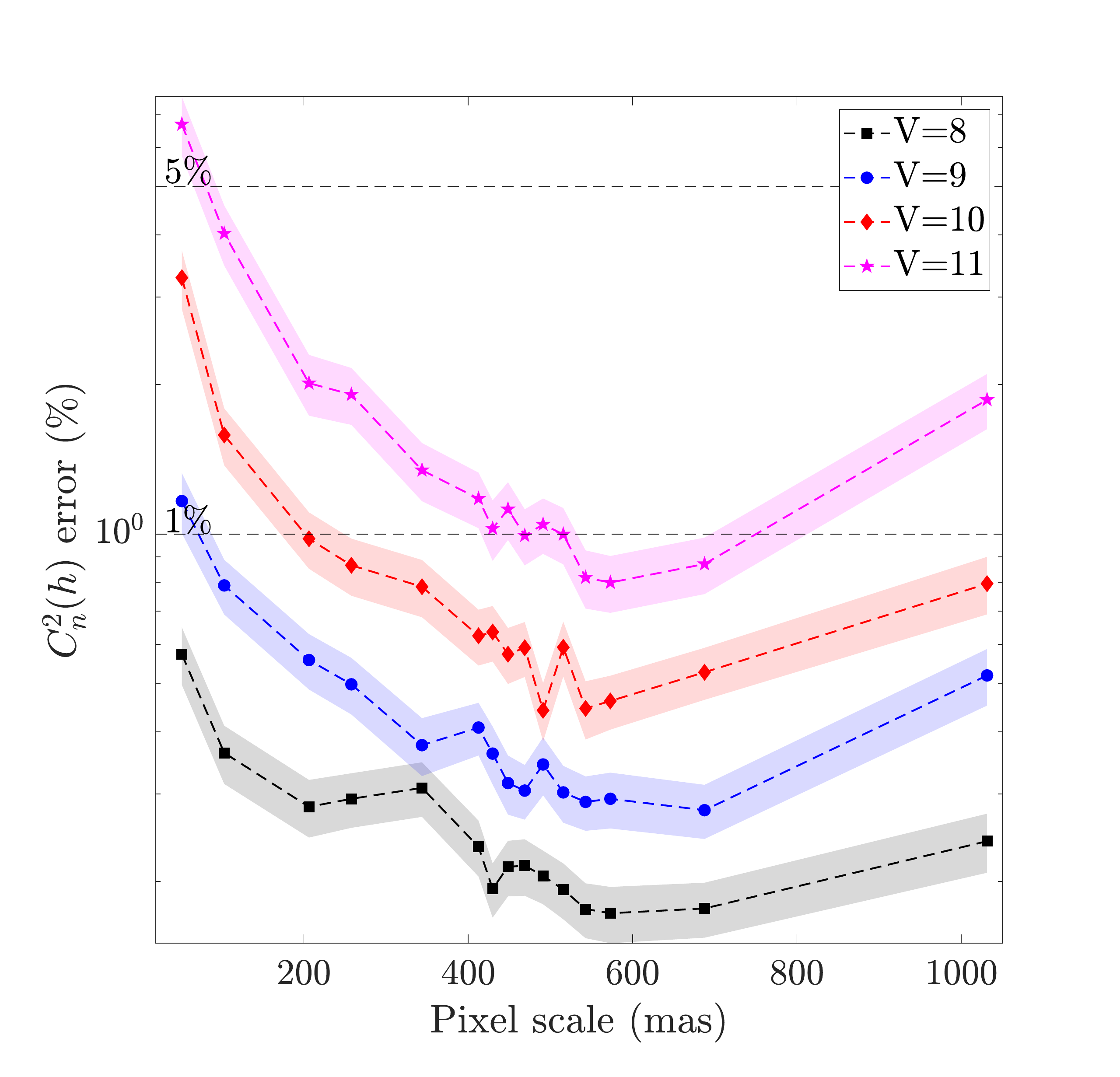}
\caption{Error on the $\cnh$ estimation regarding \textbf{Top:} the calibration star magnitude for different classes of telescope sizes and a 500mas PSF sampling \textbf{Bottom:} the pixel scale for a 0.5m telescope and different calibration star magnitude. Envelopes give the 1-$\sigma$ error bar.}
\label{F:cn2hErrorVMag}
\end{figure}

We have measured the $\cnh$ error with respect to the star magnitude for different telescope sizes and detector pixel scales as illustrated in Fig.~\ref{F:cn2hErrorVMag}. The telescope FOV was set up to the value given by Eq.~\ref{E:thetaopt} with the lowest layer at 100\,m at zenith. We notice that 1\% of accuracy can be reached on a 0.5\,m telescope by relying on a star with a magnitude V=9 up to 11, respectively with a pixel scale of 100\,mas and 500\,mas. Also, doubling the telescope diameter allows to reach the same accuracy with a star 1.5\,mag fainter.

One may work with even bigger pixels, but it would diminish the number of meaningful pixels involved in the best-fitting process, decrease the sensitivity to anisokinetism and degrade the estimation and we see in Fig.~\ref{F:cn2hErrorVMag}. The pixel scale must be settled regarding both the accessible field of view and the star magnitude. If a larger telescope is used to increase the magnitude limit, then PEPITO requires a larger field of view to reach the same minimum altitude, so this is the trade-off to be considered. With a standard 1024$\times$1024 detector and a pixel scale set up to 500\,mas, we get a field of view of 8.5\,arcmin which makes PEPITO capable of retrieving the profile above 170\,m.

\textcolor{black}{Also, results presented in Fig.~\ref{F:cn2hErrorVMag} were obtained with a narrow filter ($\delta \lambda$ = 10\,nm). When considering a 300\,nm-wide filter with a constant transmission, 1\% of error on the $\cnh$ is enabled with V=15\,mag stars and a 500\,mas sized pizel for a 0.5\,m telescope diameter. The exact hardware implementation of PEPITO is still under designing, but should certainly rely on a low-noise camera with large spectral transmission in order to increase as much as possible the S/N, in a way we can expect to have PEPITO capable of relying on V=11-15 mag stars.}

\section{Conclusions}

We have presented PEPITO as a new concept of atmospheric turbulence profiling based on wide-field, short-exposure images in seeing-limited mode. PEPITO performs a uniform TT-correction across the field by measuring the TT in a specific position in the field, which produces an anisokinetism effect that changes the PSF morphology with $\cnh$ and $L_0(h)$ profiles.
Therefore, the estimation relies on a PSF fitting process that exploits the PSF morphology to deduce the atmospheric profiles. PEPITO is operable as a built-in instrument and requires a simple optical setup with a telescope an a large detector camera. The advantage of PEPITO is that it allows to calibrate the anisoplanatism model that is served to post-process large-field SCAO-compensated images which does not dispose of information to retrieve the profiles. On top of that, the $\cnh$ and $L_0(h)$ estimation benefits a full aperture gain that enhances the S/N compared to a pupil-segmented approach. On top of that, handling PSF allows to encode the atmosphere statistics into the PSF morphology and particularly into the brightest pixels. Each pixel intensity contains a substantial information about the atmosphere distribution, that particularly increases the sensitivity to $L_0(h)$ when accumulating several PSFs in the field.\\

We have demonstrated in simulation that the on-axis TT correction of short-exposure images is well described by the anisokinetism model on a 0.5\,m and 1\,m telescope with a residual model error lower than 0.2\%. PEPITO estimates of $\cnh$ and $L_0(h)$ at respectively 0.1\% and 1\%-level of accuracy with a temporal resolution of 30\,s. To characterize a layer at height $h$, we need a FOV given by $\text{FOV}\cro{\text{arcmin}} = 2.9\pm 0.2\times \text{D}\cro{\text{m}}/\text{h}\cro{\text{km}}$ and a PSF located at the edge of this field within $\pm \theta_a$, the anisokinetic angle, which therefore permits PEPITO to retrieve several close layers from a single PSF. We have determined that the altitude resolution is $\leq400$\,m, it evolves quadratically with respect to the height and linearly with the detector pixel scale. Finally, we have pinpointed that PEPITO can obtain 1\% of accuracy in $\cnh$ from stars of magnitude V=9 up to 11 with respective pixel scales of 100\,mas and 500\,mas, using 3000 frames of 10\,ms exposure time and a detector read-out noise of 0.8\,e$^{-1}$.

In the future, we want to validate PEPITO on-sky and confirm that its outputs can serve an accurate description of the anisoplanatism for SCAO-assisted observations. In particular, we aim to enhancing the image post-processing for deconvolution of extended objects in anisoplanatic regime and model-fitting for very crowded globular clusters characterization for which the $\cnh$ can not be necessarily estimated directly from the image.

\section*{Acknowledgments}
	This work was supported by the A*MIDEX project (no. ANR-11-IDEX-0001-02) funded by the "Investissements d'Avenir" French Government programme, managed by the French National Research Agency (ANR). This work is also supported OPTICON H2020 WP10. NAB acknowledges UKRI STFC funding (ST/P000541/1).

\bibliographystyle{mnras} 
\bibliography{biblioLolo}

\bsp
\label{lastpage}	
\end{document}